\documentclass[preprint, 12pt]{aastex}
\usepackage{graphicx,natbib,amsmath,multirow}

\shorttitle{A Double-Ring Algorithm for Modeling Solar Active Regions: Unifying Kinematic Dynamo Models and Surface Flux-Transport Simulations} \shortauthors{Mu\~noz-Jaramillo, Nandy, Martens and Yeates}

\begin{document}

\title{A Double-Ring Algorithm for Modeling Solar Active Regions: Unifying Kinematic Dynamo Models and Surface Flux-Transport Simulations}

\author{Andr\'es Mu\~noz-Jaramillo}
\affil{Department of Physics, Montana State University, Bozeman, MT 59717, USA}
\email{munoz@solar.physics.montana.edu}

\and

\author{Dibyendu Nandy}
\affil{Department of Physical Sciences, Indian Institute for Science Education and Research, Kolkata, Mohampur 741252, West Bengal, India}
\email{dnandi@iiserkol.ac.in}

\and

\author{Petrus C. H. Martens}
\affil{Department of Physics, Montana State University, Bozeman, MT 59717, USA}
\affil{Harvard-Smithsonian Center for Astrophysics, Cambridge, MA 02138, USA}
%\email{pmartens@cfa.harvard.edu}
\email{martens@solar.physics.montana.edu}

\author{Anthony R. Yeates}
\affil{Division of Mathematics, University of Dundee, Dundee DD1 4HN, Scotland, UK}
\email{anthony@maths.dundee.ac.uk}

\begin{abstract}
The emergence of tilted bipolar active regions (ARs) and the dispersal of their flux, mediated via processes such as diffusion, differential rotation and meridional circulation is believed to be responsible for the reversal of the Sun's polar field.   This process (commonly known as the Babcock-Leighton mechanism) is usually modeled as a near-surface, spatially distributed $\alpha$-effect in kinematic mean-field dynamo models.  However, this formulation leads to a relationship between polar field strength and meridional flow speed which is opposite to that suggested by physical insight and predicted by surface flux-transport simulations.  With this in mind, we present an improved double-ring algorithm for modeling the Babcock-Leighton mechanism based on AR eruption, within the framework of an axisymmetric dynamo model. Using surface flux-transport simulations we first show that an axisymmetric formulation -- which is usually invoked in kinematic dynamo models -- can reasonably approximate the surface flux dynamics.  Finally, we demonstrate that our treatment of the Babcock-Leighton mechanism through double-ring eruption leads to an inverse relationship between polar field strength and meridional flow speed as expected, reconciling the discrepancy between surface flux-transport simulations and kinematic dynamo models.

%The main mechanism for the recreation of poloidal field is believed to be the emergence of Active Regions, and their subsequent diffusion and transport towards the poles.  This mechanism (also known as Babcock-Leighton mechanism) has been modeled successfully by surface flux-transport simulations of increasing sophistication during the last two decades.  As opposed to surface flux transport simulations where the system is driven by active region bipolar pairs, in kinematic dynamo models the Babcock-Leighton mechanism is modeled through a mean-field formulation.  However, after the initial introduction of this formulation little has been done to improve upon it.  Here we further the idea of using axisymmetric ring doublets to model individual active regions and thus bring the implementation of the Babcock-Leighton mechanism in dynamo models closer to its roots.

%In order to evaluate the validity of an axisymmetric representation of active regions we perform surface flux transport simulations whose results imply that for time-scales relevant to the solar cycle the axisymmetric formulation is adequate.  Additionally, by comparing the performance of the mean-field formulation versus the double-ring algorithm, we show that the later captures the surface dynamics better than mean-field formulation and resolves previously found discrepancies between kinematic dynamo models and surface flux transport simulations.
\end{abstract}


\section{Introduction}

Currently, some of the best tools for understanding the solar magnetic cycle are axisymmetric kinematic dynamo models and surface flux-transport simulations.   On the one hand kinematic dynamo models (which are usually based on an axisymmetric formulation), attempt to model the magnetic cycle self-consistently by using a prescribed meridional flow, differential rotation, turbulent diffusivity and poloidal source (see Sec.~\ref{Sec_KMFD}).  They have been successful in reproducing several of the characteristics of the solar cycle (see for example: Choudhuri, Sch\"ussler \& Dikpati 1995\nocite{choudhuri-schussler-dikpati95}; Durney 1997\nocite{durney97}; Dikpati \& Charbonneau 1999\nocite{dikpati-charbonneau99}; Covas et al.~2000\nocite{covas-etal00}; Nandy \& Choudhuri 2001\nocite{nandy-choudhuri01}; Rempel 2006\nocite{rempel06}; Guerrero \& de Gouveia Dal Pino 2007\nocite{guerrero-degouveiadalpino07}; Jouve \& Brun 2007\nocite{jouve-brun07}; Mu\~noz-Jaramillo, Nandy \& Martens 2009\nocite{munoz-nandy-martens09}, MNM09 from here on; for more information about kinematic dynamo models see review by Charbonneau 2005\nocite{charbonneau05}).  On the other hand, surface flux-transport simulations study the evolution of the photospheric magnetic field by integrating the induction equation using a prescribed meridional flow, differential rotation and turbulent diffusivity.  There are two main differences between surface flux-transport simulations and kinematic dynamo models: in the former the computational domain is restricted to the surface (without imposing axisymmetry) and they are not self-excited, but driven by the deposition of active region (AR) bipolar pairs.  This type of models has proved a successful tool for understanding surface dynamics on long timescales (see, for example, Mackay, Priest \& Lockwood 2002\nocite{mackay-priest-lockwood02b}; Wang, Lean \& Sheeley 2002\nocite{wang-lean-sheeley02}; Schrijver, De Rosa \& Title 2002\nocite{schrijver-derosa-title02}) and the evolution of coronal and interplanetary magnetic field (see for example Lean, Wang \& Sheeley 2002\nocite{lean-wang-sheeley02}; Yeates, Mackay \& van Ballegooijen 2008\nocite{yeates-mackay-vanballegooijen08}).  However, a discrepancy exists between kinematic dynamo models and surface flux-transport simulations regarding the relationship between meridional flow amplitude and the strength of the polar field (Schrijver \& Liu 2008\nocite{schrijver-liu08}; Hathaway \& Rightmire 2010\nocite{hathaway-rightmire10}; Jiang et al. 2010\nocite{jiang-etal10}).  On the one hand kinematic dynamo models find that a stronger meridional flow results in stronger polar field (Dikpati, de Toma \& Gilman 2008\nocite{dikpati-detoma-gilman08}), on the other hand surface flux-transport simulations find an inverse relationship (Wang, Sheeley \& Lean 2002\nocite{wang-sheeley-lean02}; Jiang et al. 2010\nocite{jiang-etal10}).   In this work we improve upon the idea proposed by Durney (1997\nocite{durney97}) and further elucidated by Nandy \& Choudhuri (2001\nocite{nandy-choudhuri01}) of using axisymmetric ring doublets to model individual ARs.  We show that this captures the surface dynamics better than the $\alpha$-effect formulation and resolves the discrepancy between dynamo models and surface flux-transport simulations regarding the relationship between meridional flow speed and polar field strength.

\section{Evolution of the Axisymmetric Component of the Magnetic Field on Timescales Comparable to the Solar Cycle}\label{Sec_FTS_CH3}

\begin{figure}
\begin{center}
\begin{tabular}{cc}
                \textbf{Case 1 }                &                   \textbf{Case 2} \\
  \includegraphics[scale=0.5]{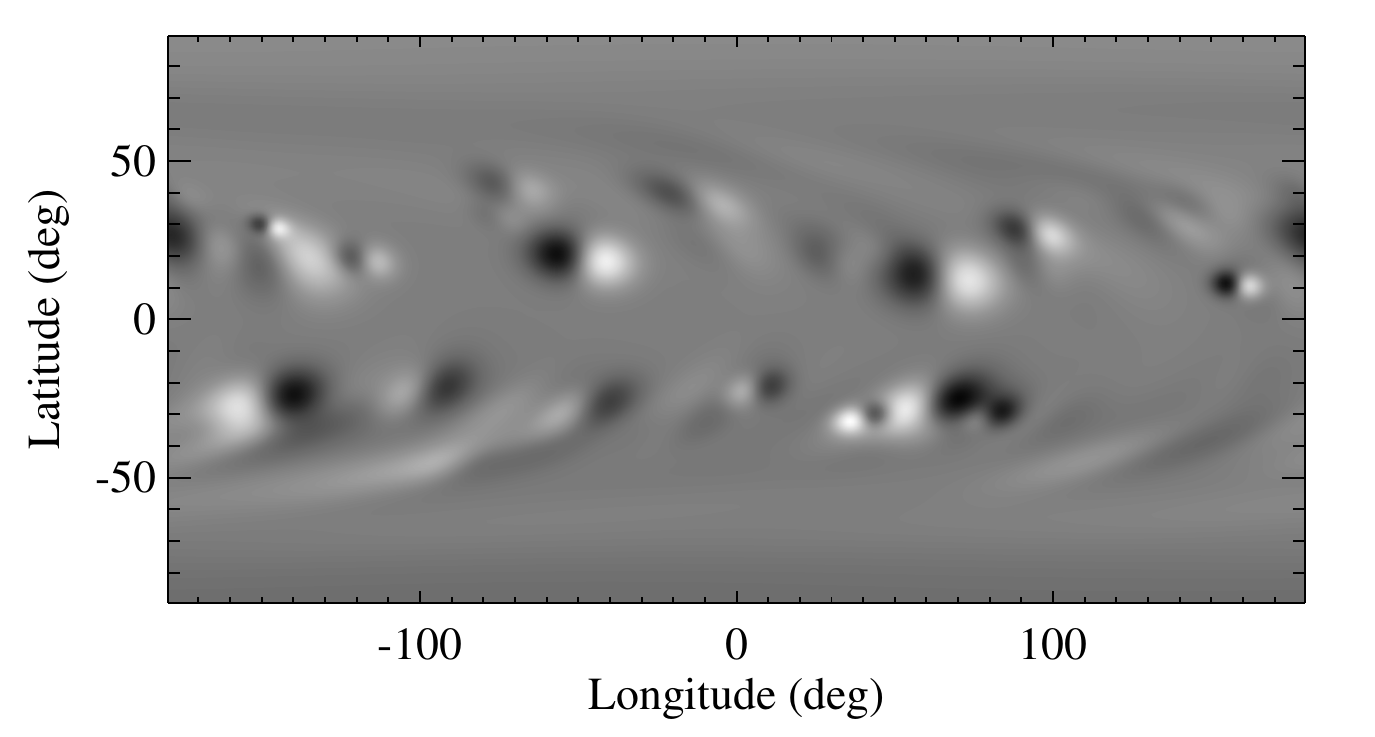} & \includegraphics[scale=0.5]{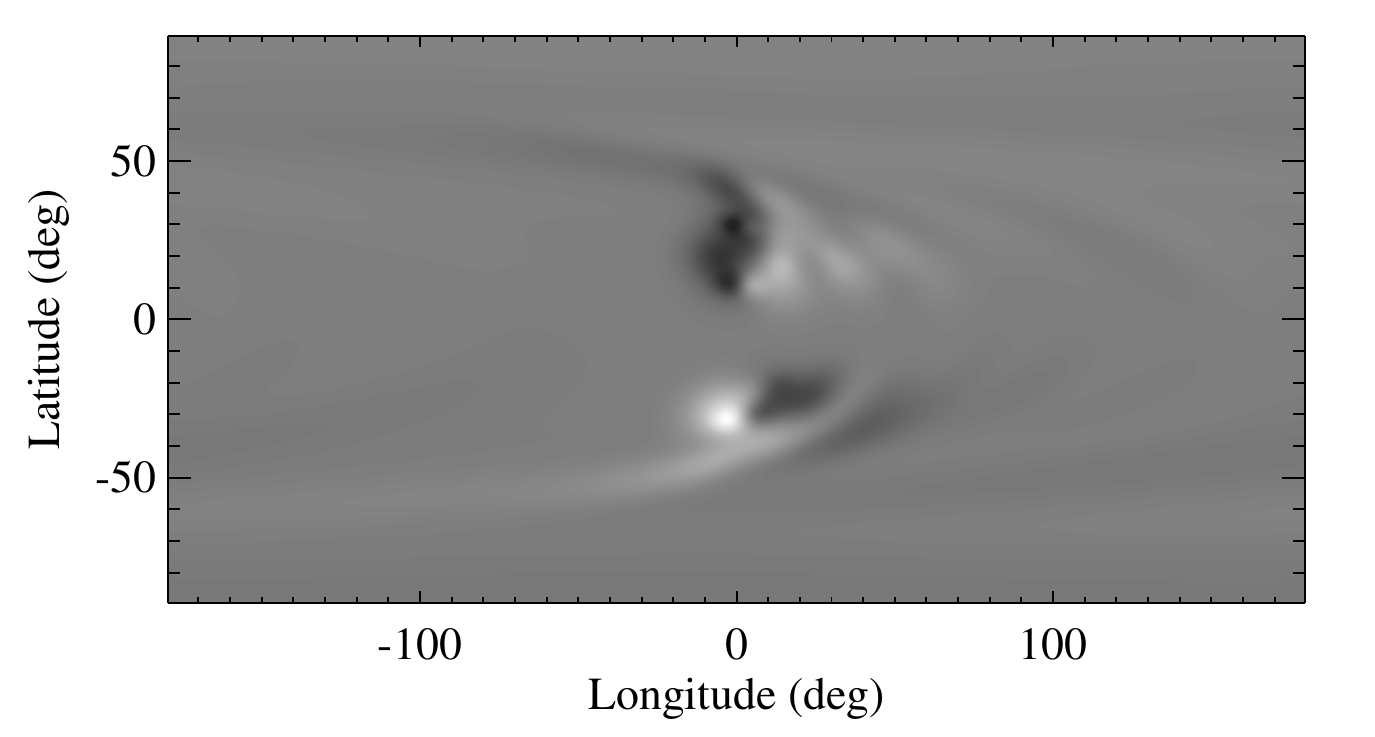}\\
                 (a)                   &                   (b)\\
  \includegraphics[scale=0.35]{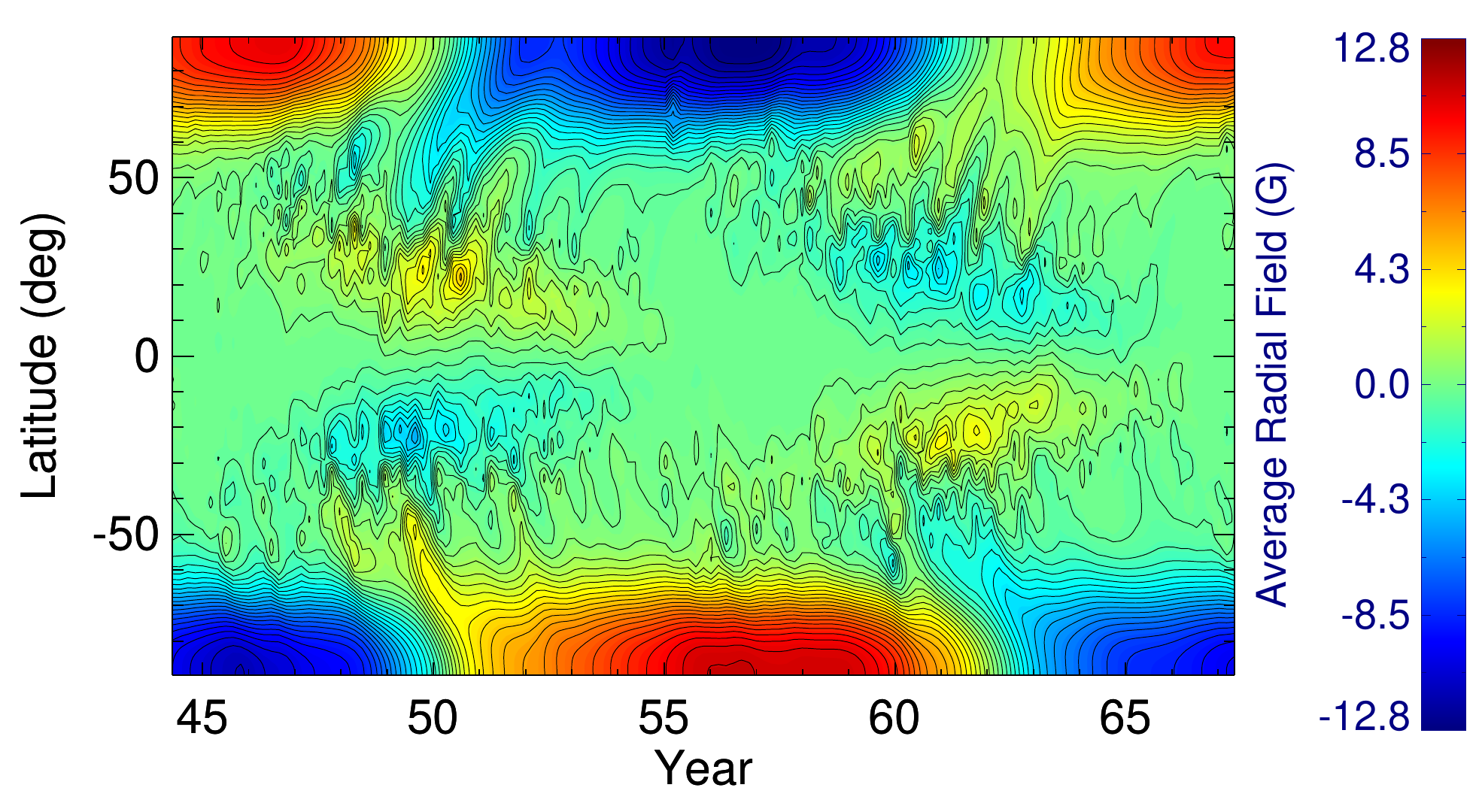} & \includegraphics[scale=0.35]{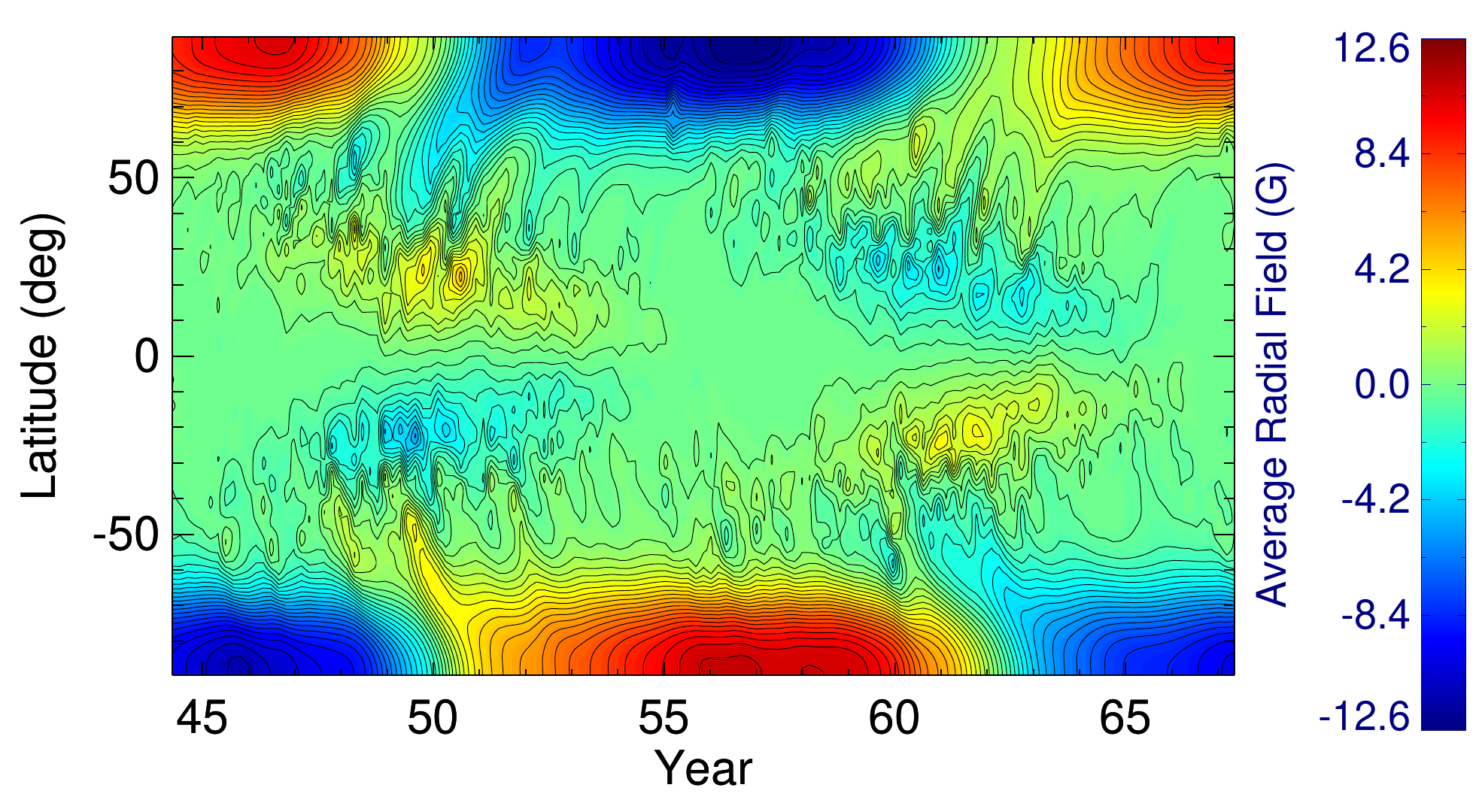}\\
                 (c)                   &                   (d)\\\\
                 \hline
  \includegraphics[scale=0.35]{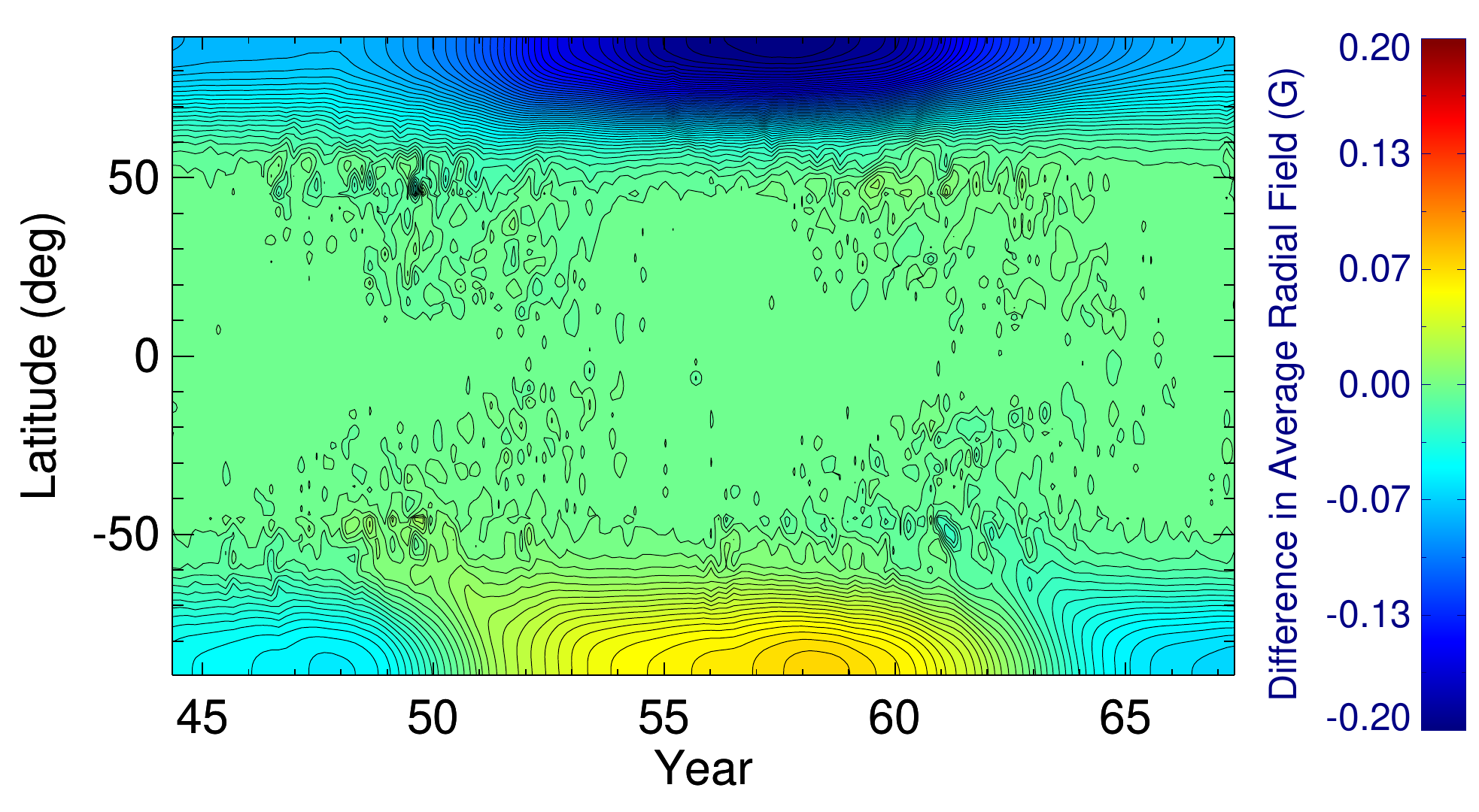} & \includegraphics[scale=0.5]{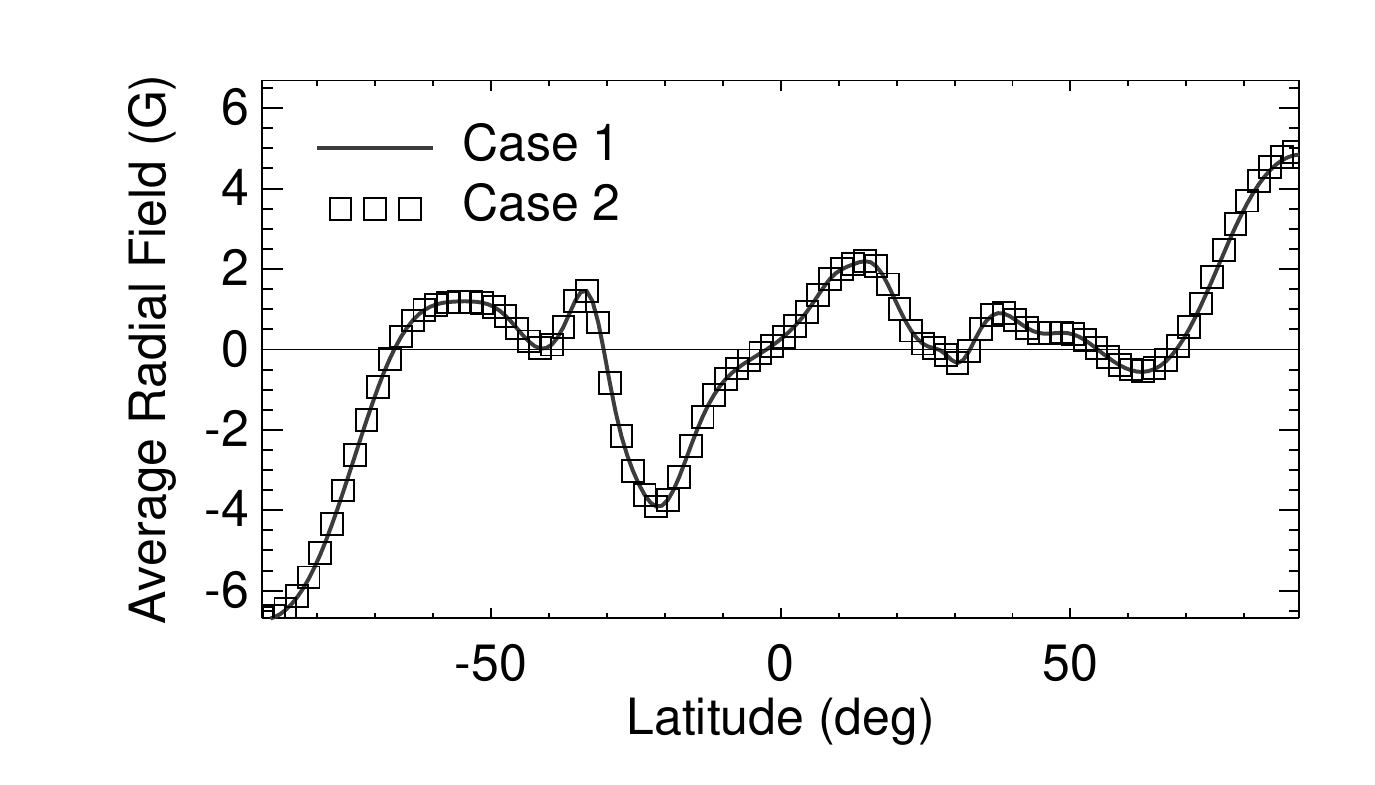}\\
                 (e)                   &                   (f)
\end{tabular}
\end{center}
\caption{Long term evolution of the photospheric magnetic field in surface flux-transport simulations: (a) Snapshot of the magnetic field at the peak of the cycle for Case 1 (active regions deposited at all longitudes). (b) Snapshot of the magnetic field at the peak of the cycle for Case 2 (active regions deposited at a single longitude).  (c) Butterfly diagram for Case 1.  (d) Butterfly diagram for Case 2.  (e) Difference between the butterfly diagrams of Case 1 and Case 2.  (f) Longitudinal average of the snapshots shown in the top row, taken at the peak of the cycle.}\label{Fig_SFT}
\end{figure}

As mentioned before, kinematic dynamo models are usually based on an axisymmetric formulation and our model is not an exception.  Given that here we introduce an improved axisymmetric double-ring algorithm for modeling AR eruptions (see below), but AR emergence is strictly a non-axisymmetric process, it is important to study the amount of information lost by averaging over the longitudinal dimension.  We do this by performing surface transport simulations driven by a synthetic set of AR cycles based on Kitt Peak data using the model of Yeates, Mackay \& van Ballegooijen (2007\nocite{yeates-mackay-vanballegooijen07}).  We perform a regular surface flux-transport simulation in which the bipolar ARs are distributed all across the surface of the Sun (Case 1) and another in which the same set of ARs is deposited at the same Carrington longitude while leaving other properties (time, tilt, latitude of emergence and flux) intact (Case 2).  The difference between both simulations is clear from the top row of Fig.~\ref{Fig_SFT}, where we show a snapshot of the surface magnetic field at the peak of the cycle for Case 1 (Fig.~\ref{Fig_SFT}-a) and Case 2 (Fig.~\ref{Fig_SFT}-b).  Obviously these cases have entirely different magnetic configurations at the time of deposition.  However, when the magnetic field is averaged in longitude and stacked in time to create a magnetic synoptic map (also know as butterfly diagram; Figs.~\ref{Fig_SFT}-c \& \ref{Fig_SFT}-d), a careful examination shows that the results are essentially the same within a margin of 1\% (Figs.~\ref{Fig_SFT}-e \& f).  The reason the simulations have identical outcomes is that the differential rotation and the meridional flow are both independent of longitude in the simulations.  Note that non-axisymmetry is essential for the evolution of the corona and interplanetary magnetic field.  This result simply indicates that an axisymmetric representation of surface dynamics is a reasonable approximation if we are only concerned with the general properties of the magnetic field at the surface over solar cycle timescales in the context of dynamo models.

\section{Modeling Individual Active Regions as Axisymmetric Double-Rings}\label{Sec_AR}

\begin{figure}
\begin{center}
\begin{tabular}{cc}
  \includegraphics[scale=0.45]{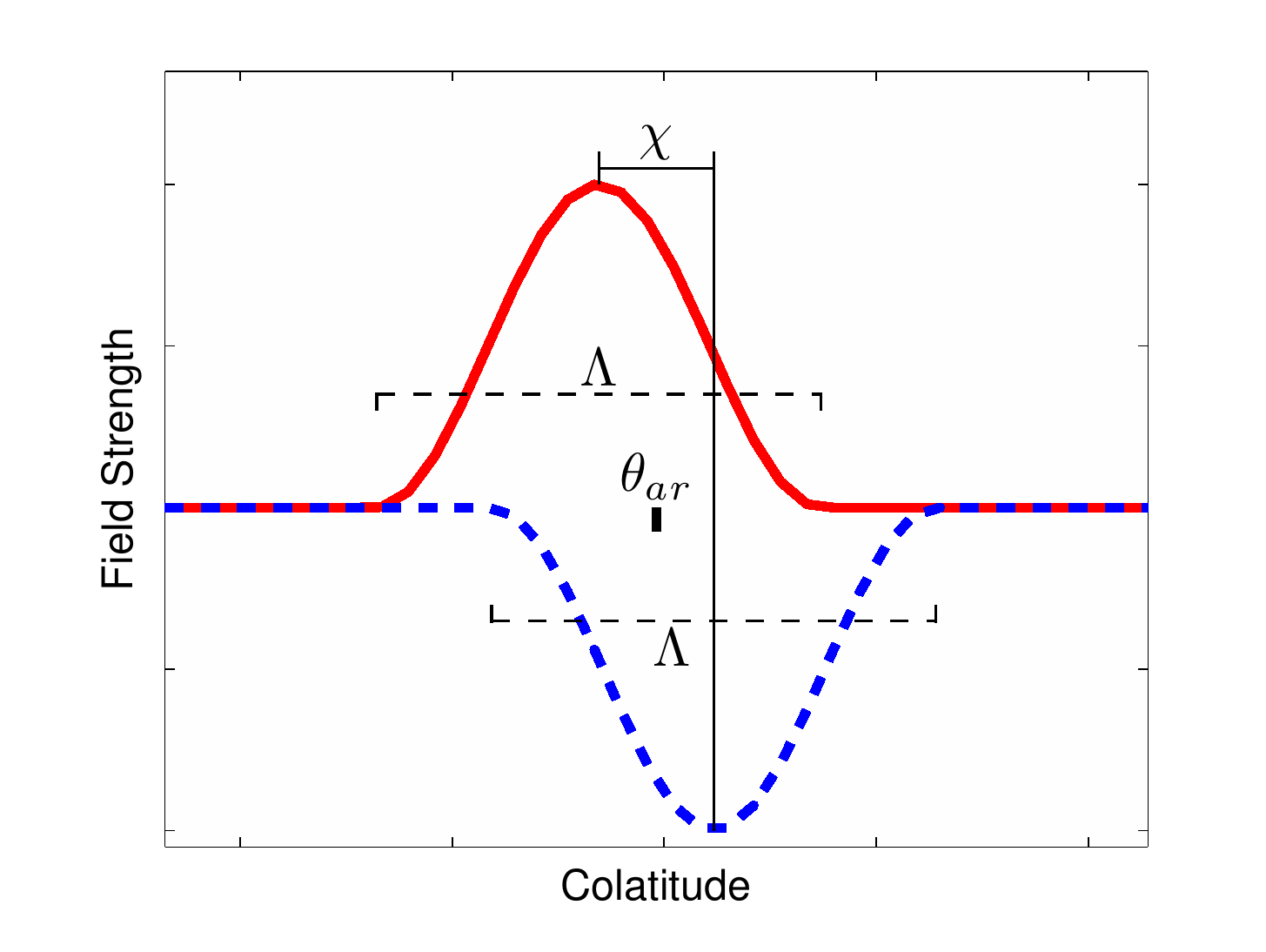} & \includegraphics[scale=0.45]{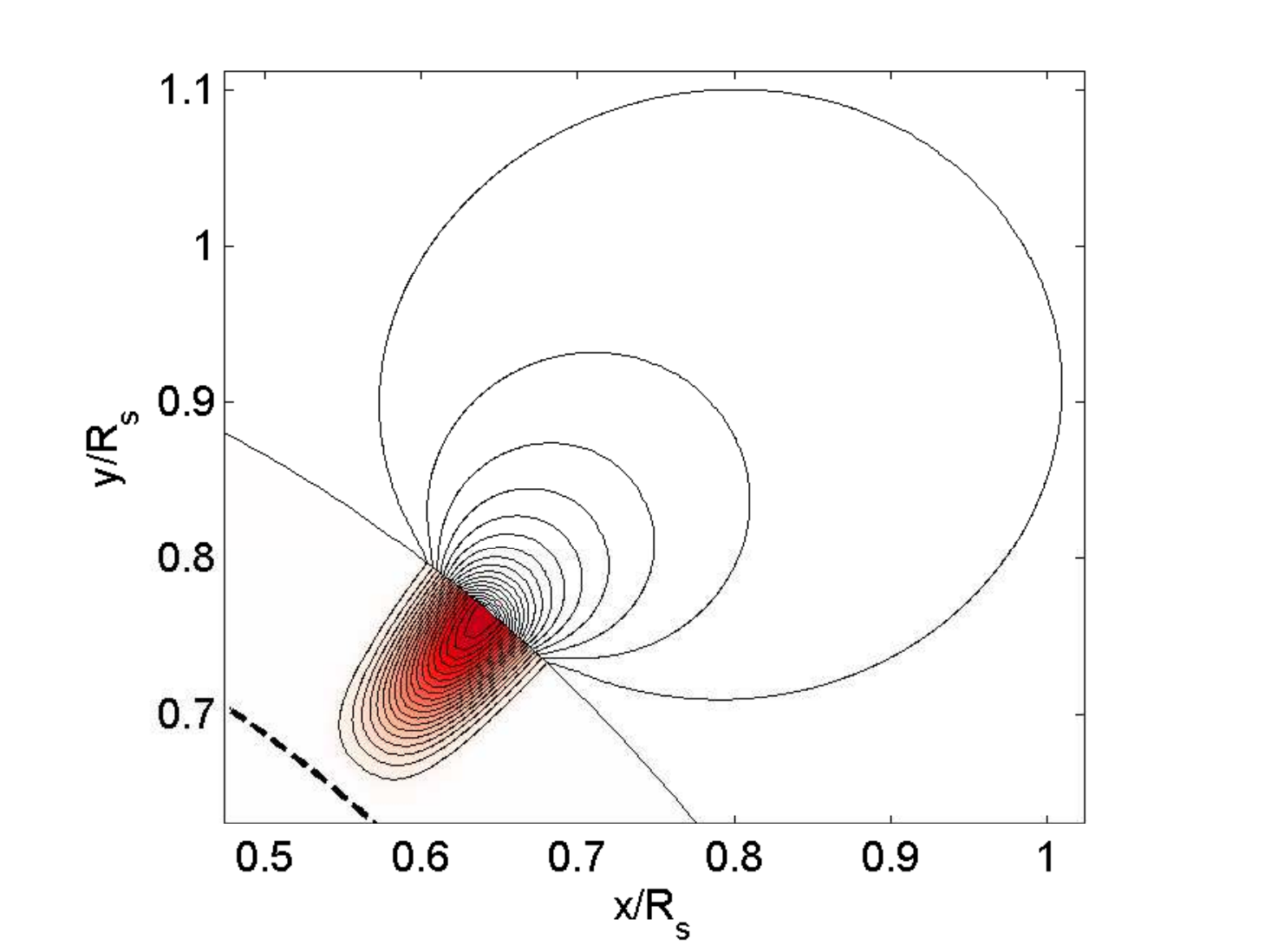}\\
                 (a)                   &                   (b)
\end{tabular}
\end{center}
\caption{(a) Diagram illustrating the quantities which define the latitudinal dependence of a double-ring bipolar pair.  (b) Poloidal field lines of one of our double-rings including a potential field extrapolation for the region outside the Sun.  The dashed line marks the location of the penetration depth $R_{ar}$.}\label{Fig_DR}
\end{figure}

The initial implementation of the double-ring algorithm by Durney (1997\nocite{durney97}) and Nandy \& Choudhuri (2001\nocite{nandy-choudhuri01}) consisted in searching the bottom of the convection zone (CZ) for places in which the toroidal field exceeds a buoyant threshold and placing two axisymmetric rings of constant radial flux directly above them.  This implementation had two important deficiencies: strong sensitivity to changes in grid resolution and the introduction of sharp discontinuities in the $\phi$ component of the vector potential.  The first necessary step to address these problems is a careful mathematical definition of the vector potential associated with each ring doublet, which ensures a continuous first derivative in the computational domain.  We do so by building a separable function:
\begin{equation}\label{Eq_AR}
    A_{ar}(r,\theta)= K_0 A(\Phi)F(r)G(\theta),
\end{equation}
where $K_0$ is a constant we introduce to ensure super-critical solutions and $A(\Phi)$ defines the strength of the ring doublet. $F(r)$ is defined as
\begin{equation}\label{Eq_AR_R}
    F(r)= \left\{\begin{array}{cc}
            0 & r<R_\odot-R_{ar}\\
            \frac{1}{r}\sin^2\left[\frac{\pi}{2 R_{ar}}(r - (R_\odot-R_{ar}))\right] & r\geq R_\odot-R_{ar}
          \end{array}\right.,
\end{equation}
where $R_\odot=6.96\times10^8$ m corresponds to the radius of the Sun and $R_{ar}=0.85R_\odot$ represents the penetration depth of the AR.  This depth is motivated from results indicating that the disconnection of an AR flux-tube happens deep down in the CZ (Longcope \& Choudhuri 2002\nocite{longcope-choudhuri02}).  $G(\theta)$, on the other hand, is easier to define in integral form:
\begin{equation}\label{Eq_AR_Th}
    G(\theta) = \frac{1}{\sin{\theta}}\int_0^{\theta}[B_{-}(\theta')+B_{+}(\theta')]\sin(\theta')d\theta',
\end{equation}
where $B_{+}$ ($B_{-}$) defines the positive (negative) ring:
 \begin{equation}\label{Eq_AR_Dp}
    B_{\pm}(\theta)= \left\{\begin{array}{cc}
                     0 & \theta<\theta_{ar}\mp\frac{\chi}{2}-\frac{\Lambda}{2}\\
                     \pm\frac{1}{\sin(\theta)}\left[1+\cos\left(\frac{2\pi}{\Lambda}(\theta-\theta_{ar}\pm\frac{\chi}{2})\right)\right] & \theta_{ar}\mp\frac{\chi}{2}-\frac{\Lambda}{2} \leq \theta < \theta_{ar}\mp\frac{\chi}{2}+\frac{\Lambda}{2}\\
                     0 & \theta \geq \theta_{ar}\mp\frac{\chi}{2}+\frac{\Lambda}{2}
               \end{array}\right..
\end{equation}
Here $\theta_{ar}$ is the co-latitude of emergence, $\Lambda$ is the diameter of each polarity of the doublet , for which we use a fixed value of $6^o$ (heliocentric degrees) and $\chi = \arcsin[\sin(\gamma)\sin(\Delta_{ar})]$ is the latitudinal distance between the centers, which in turn depends on the angular distance between polarity centers $\Delta_{ar}=6^o$ and the AR tilt angle $\gamma$; $\chi$ is calculated using the spherical law of sines (see Fig.~\ref{Fig_DR}-a for a diagram illustrating these quantities).  Fig.~\ref{Fig_DR}-b shows the axisymmetric signature of one of such axisymmetric ARs.

\section{Recreating the Poloidal Field}\label{sec_dblr2pld}

Given that the accumulated effect of all ARs is what regenerates the poloidal field, we need to specify an algorithm for AR eruption and decay in the context of the solar cycle.  On each solar day of our simulation we randomly chose one of the latitudes with fields higher than a buoyancy threshold of $B_c = 5\times 10^4$ Gauss at the bottom of the CZ ($r=0.71R_\odot$), and calculate the amount of magnetic flux present within its associated toroidal ring. The probability distribution we use is not uniform, but is restricted to observed active latitudes.  We do this by making the probability function drop steadily to zero between 30$^o$ (-30$^o$) and 40$^o$ (-40$^o$) in the northern (southern) hemisphere:
\begin{equation}\label{Eq_Prob}
    P(\theta)\propto\left( 1 + \operatorname{erf}\left[ \frac{\theta - 0.305\pi}{0.055\pi} \right] \right)\left( 1 - \operatorname{erf}\left[ \frac{\theta - 0.694\pi}{0.055\pi} \right] \right).
\end{equation}
We then calculate the corresponding AR tilt, using the local field strength $B_0$, the calculated flux $\Phi_0$ and the latitude of emergence $\lambda$.  For this we use the expression found by Fan, Fisher \& McClymont (1994\nocite{fan-fisher-mcclymont94}):
\begin{equation}\label{Eq_Prob}
   \gamma \propto \Phi_0^{1/4}B_0^{-5/4}\sin(\lambda),
\end{equation}
reducing the magnetic field of the toroidal ring from which the AR originates.  In order to do this, we first estimate how much magnetic energy is present on a partial toroidal ring (after removing a chunk with the same angular size as the emerging AR).  Given that this energy is smaller than the one calculated with a full ring, we set the value of the toroidal field such that the energy of a full toroidal ring filled with the new magnetic field strength is the same as the one calculated with the old magnitude for a partial ring.  Finally, we deposit a double-ring (as defined in Section \ref{Sec_AR}) with these calculated properties, at the chosen latitude.

\section{The Kinematic Mean-Field Dynamo Model}\label{Sec_KMFD}

We perform dynamo simulations to explore how the double-ring formulation compares to the near surface $\alpha$-effect formulation.  In particular we focus on the relationship between meridional flow speed and polar field strength.  Our model is based one the axisymmetric dynamo equations:
\begin{equation}\label{Eq_2.5DynA}
    \frac{\partial A}{\partial t} + \frac{1}{s}\left[ \textbf{v}_p \cdot \nabla (sA) \right] = \eta\left( \nabla^2 - \frac{1}{s^2}  \right)A + \alpha_0f(r,\theta)F(B_{tc})B_{tc}
\end{equation}
\begin{equation}\label{Eq_2.5DynB}
    \frac{\partial B}{\partial t}  + s\left[ \textbf{v}_p \cdot \nabla\left(\frac{B}{s} \right) \right] + (\nabla \cdot \textbf{v}_p)B = \eta\left( \nabla^2 - \frac{1}{s^2}  \right)B + s\left(\left[ \nabla \times (A\bf \hat{e}_\phi) \right]\cdot \nabla \Omega\right)   + \frac{1}{s}\frac{\partial (sB)}{\partial r}\frac{\partial \eta}{\partial r},
\end{equation}
where A is the $\phi$-component of the vector potential (from which $B_r$ and $B_\theta$ can be obtained), B is the toroidal field ($B_\phi$), $v_p$ is the meridional flow, $\Omega$ the differential rotation, $\eta$ the turbulent magnetic diffusivity and $s = r\sin(\theta)$.  The second term on the right-hand side of Equation~\ref{Eq_2.5DynA} corresponds to the poloidal source in the mean-field formulation.   In this formulation $\alpha_0$ is a constant that sets the strength of the source term and is usually used to ensure super-critical solutions; $\alpha(r,\theta)$ captures the spatial properties of the BL mechanism: confinement to the surface, observed active latitudes and latitudinal dependence of tilt, while $F(B_{tc})$ adds nonlinearity to the dynamo by quenching the source term for values of the toroidal field at the bottom of the CZ $B_{tc}$ that are too strong or too weak.  More information about this source can be found in MNM09.  Note that for simulations using the double-ring algorithm this term is not present in the equations ($\alpha_0=0$).

In order to integrate these equations, we need to prescribe four ingredients: meridional flow, differential rotation, the poloidal field regeneration mechanism, and turbulent magnetic diffusivity.  For the differential rotation, we use the analytical form of Charbonneau et al.\ (1999\nocite{charbonneau-etal99}), with a tachocline centered at $0.7R_\odot$ whose thickness is $0.05R_\odot$ and we use the meridional flow profile defined in MNM09.  This meridional flow better captures the features present in helioseismic data, specially the latitudinal dependence.  We use an amplitude of $20$ m/s for the results shown in Figure~\ref{Fig_MFvsDR} and a variable amplitude for the results shown in Figure~\ref{Fig_MF_Obs_Ch3} (see below).  We use a double stepped diffusivity profile as described in MNM09.  It starts with a diffusivity value $\eta_{bcd} = 10^8$ cm$^2$/s at the bottom of the CZ, jumps to a value of $\eta_{cz} = 10^{11}$ cm$^2$/s in the CZ, and then to a value of $\eta_{sg} = 10^{12}$ cm$^2$/s in the near-surface layers.  The first step is centered at $r_{cz} = 0.71R_\odot$ and has a half-width of $d_{cz} = 0.015R_\odot$ and the second step is centered at $r_{sg} = 0.95R_\odot$ and has a half-width of $d_{sg} = 0.025R_\odot$.

For the poloidal field regeneration mechanism we use the improved ring-doublet algorithm described above, using a value of $K_0 = 400$, in order to insure super-criticality (for a meridional flow of $30$ m/s).  For those simulations which use an $\alpha$-effect formulation, we use the non-local poloidal source described above (more information in MNM09) using a value of $\alpha_0 = 0.25$, in order to insure super-criticality (for a meridional flow of $30$ m/s).

\section{Addressing the Discrepancy Between Kinematic Dynamo Models and Surface Flux-Transport Simulations}

\begin{figure}
\begin{center}
\begin{tabular}{cc}
                     \textbf{Double-ring Algorithm}            &                   \textbf{$\alpha$-effect Formulation} \\
  \includegraphics[scale=0.5]{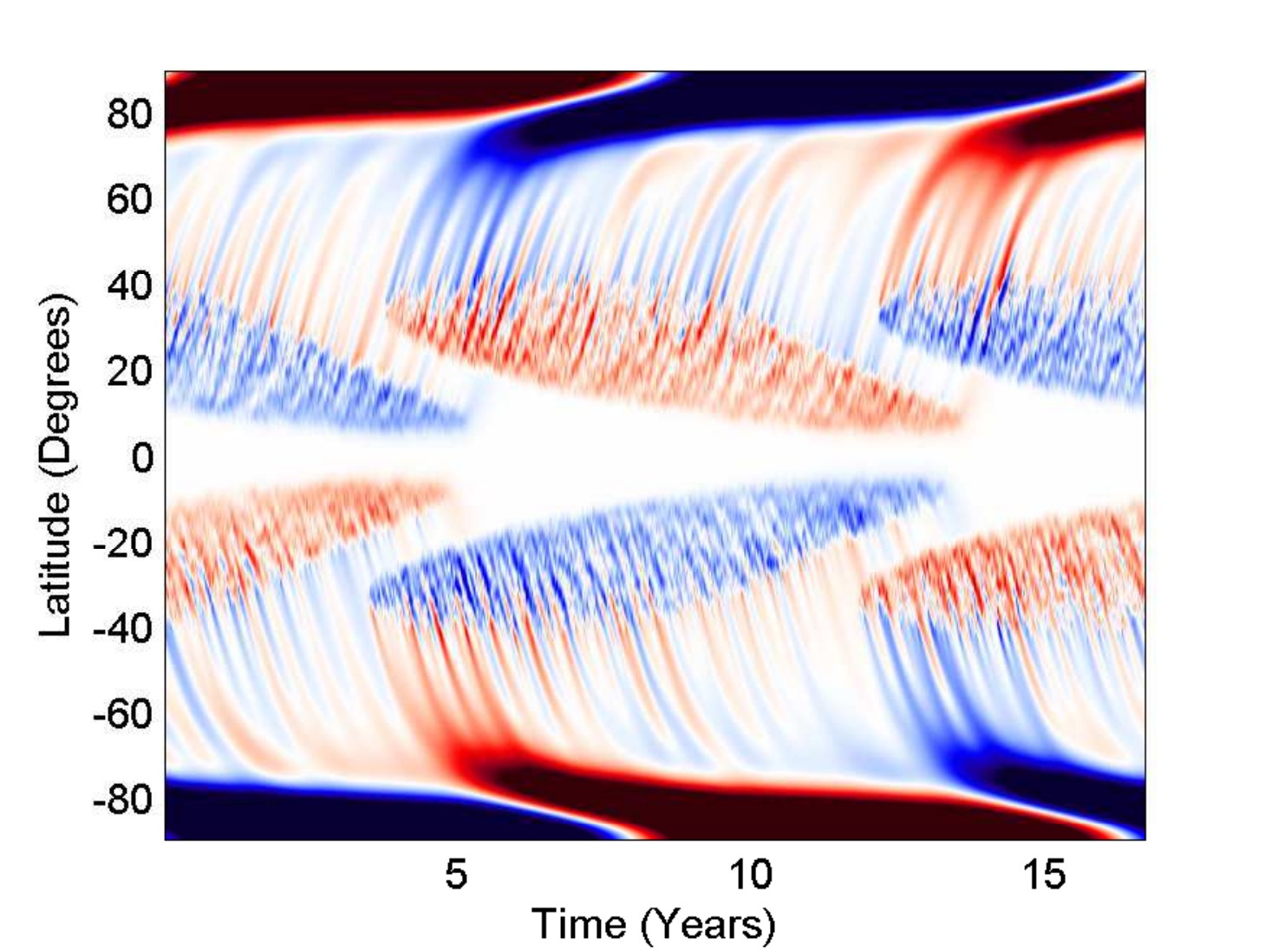} & \includegraphics[scale=0.5]{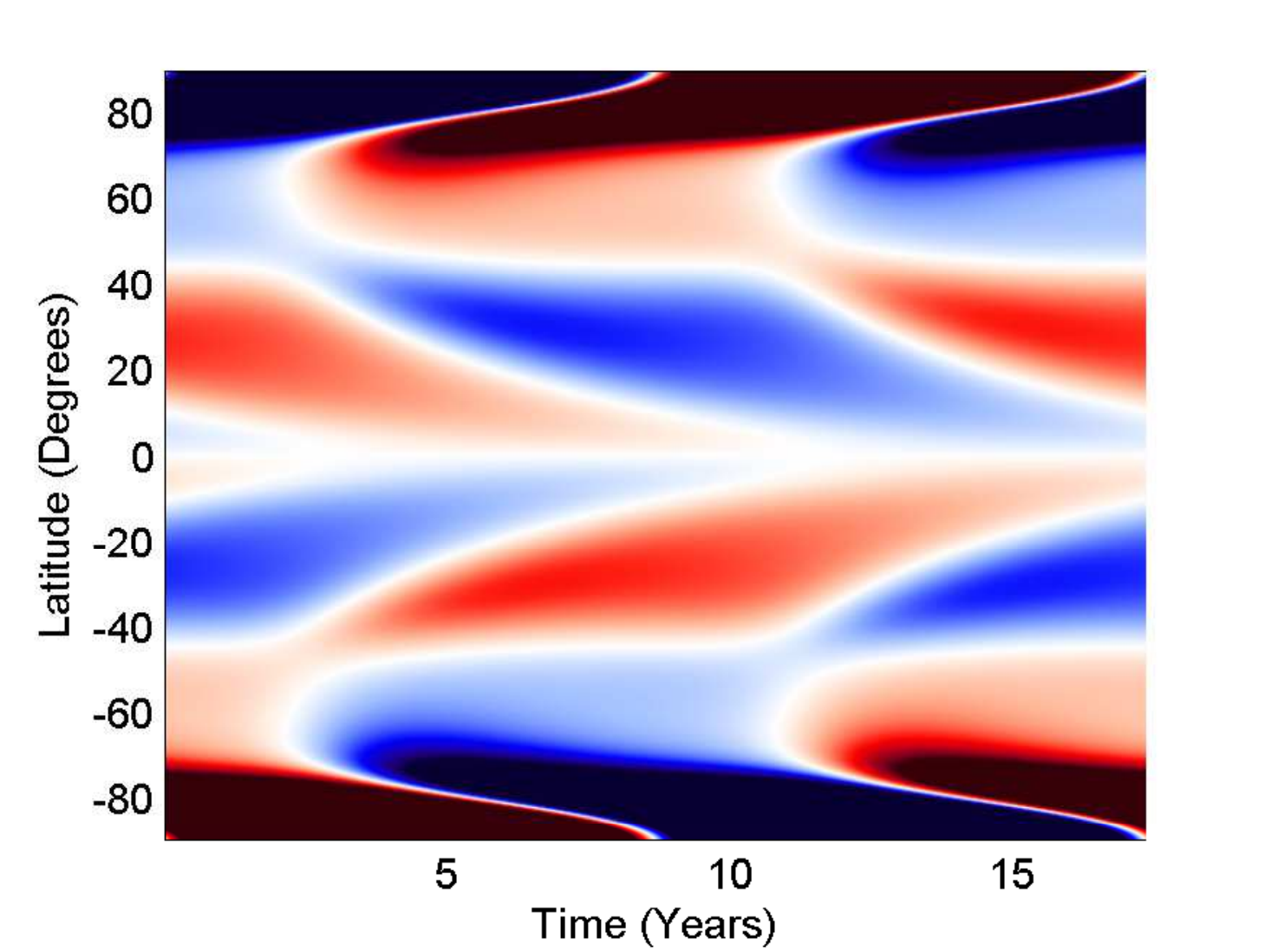}\\
                 (a)                   &                   (b)\\
  \includegraphics[scale=0.5]{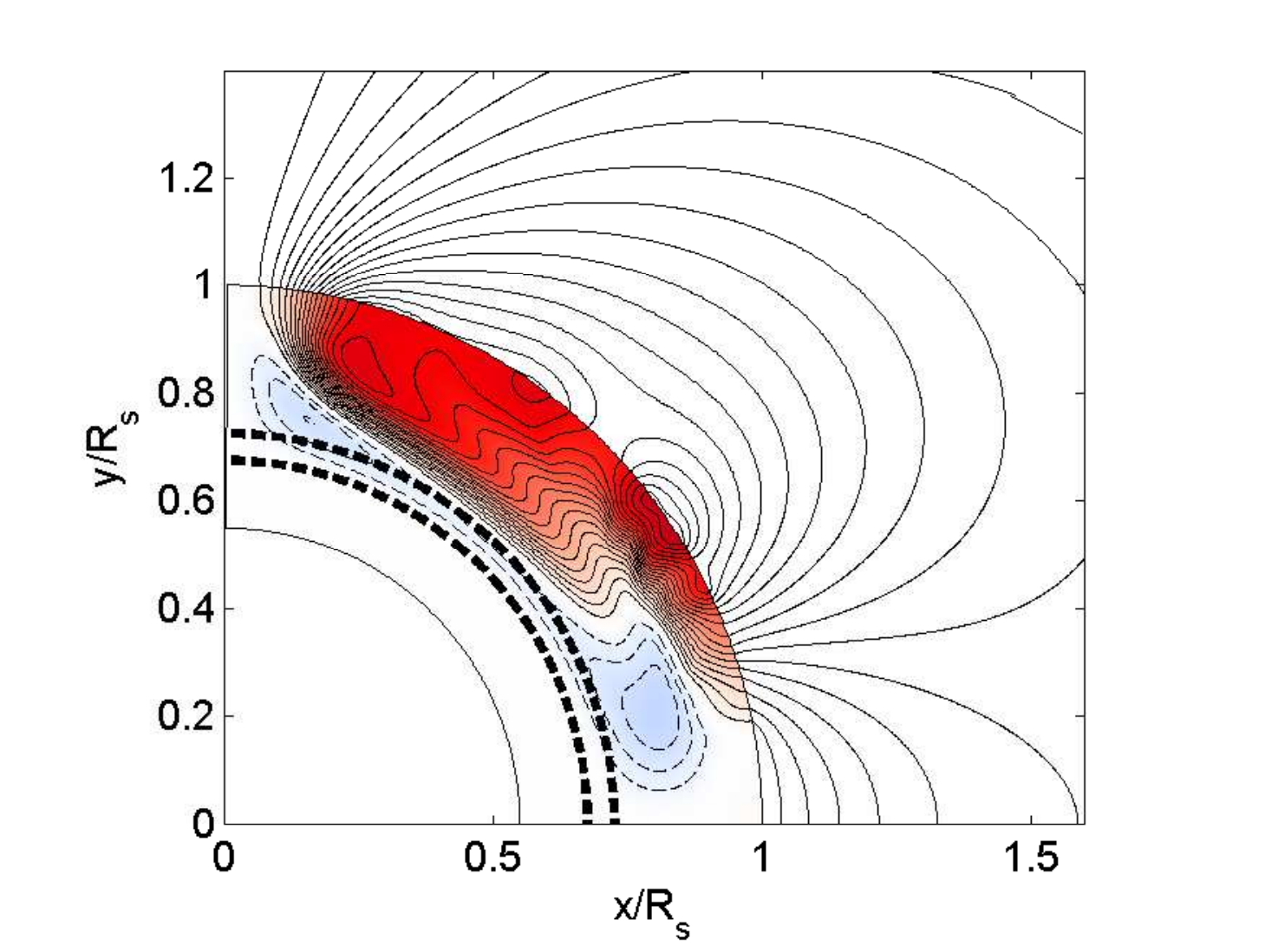} & \includegraphics[scale=0.5]{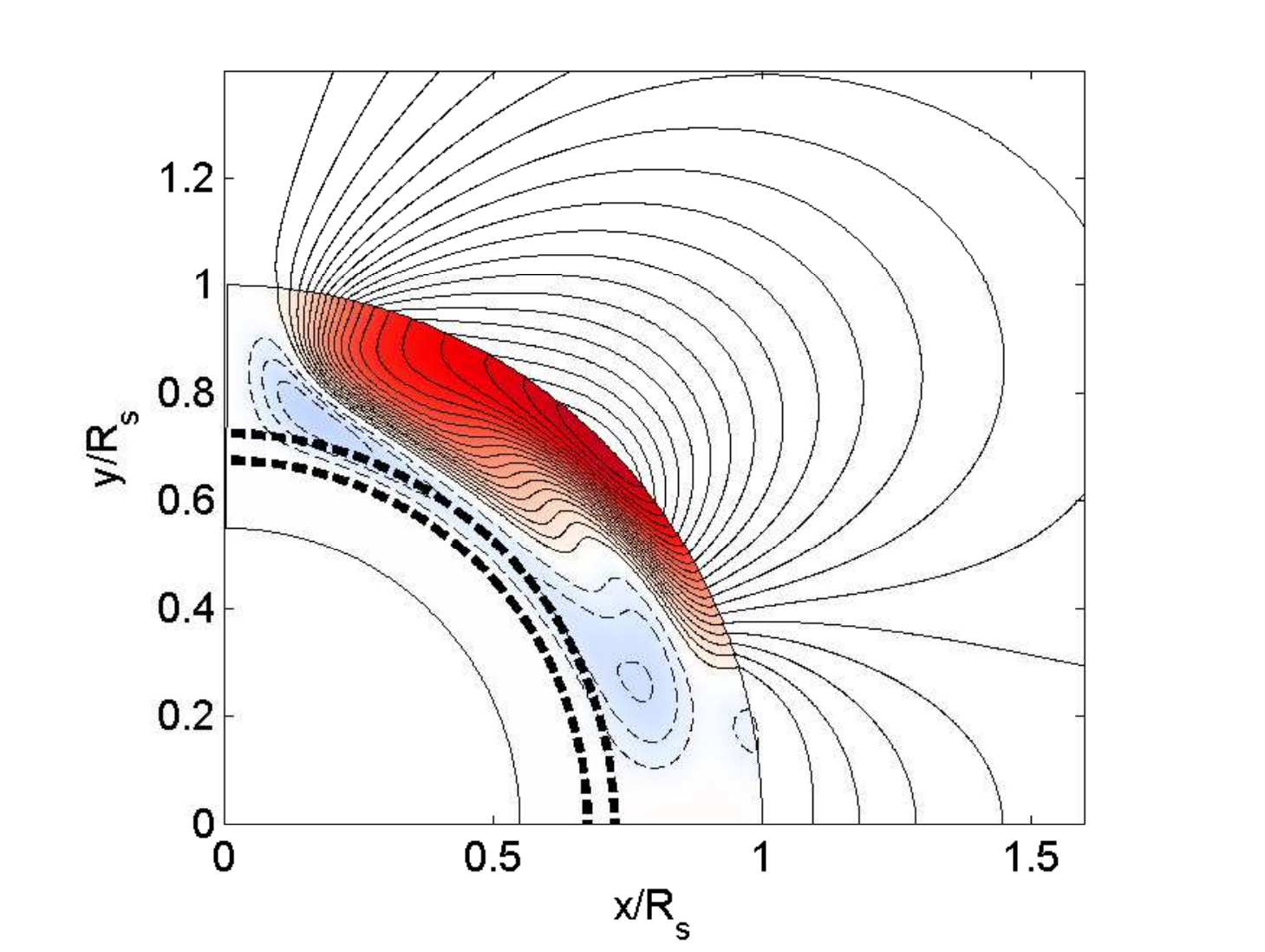}\\
                 (c)                   &                   (d)
\end{tabular}
\end{center}
\caption{Comparison between surface dynamics as captured by the double-ring algorithm (left column) and the $\alpha$-effect formulation (right column). The top row shows the evolution of the surface magnetic field in the form of synoptic maps -- the colormap is saturated to enhance the visibility of the field at mid to low latitudes.  The bottom row shows a snapshot of the poloidal components of the magnetic field taken at solar max.  The solid contours corresponds to clockwise field-lines, the dashed contours correspond to counter-clockwise field-lines.  The thick dashed lines mark the location of the tachocline.}\label{Fig_MFvsDR}
\end{figure}

In order to have a net accumulation of unipolar field at the poles, it is necessary to have an equal amount of flux cancellation across the equator.  Since the meridional flow is poleward in the top part of the convection zone, it essentially acts as a barrier against flux cancellation by sweeping both positive and negative AR polarities towards the poles resulting in weak polar fields.  This leads to an inverse correlation between flow speed and polar field strength which is accurately captured in surface flux transport simulations.   Contrarily, dynamo simulations in typically used parameter regimes obtain an opposite relationship not consistent with the above physics.   This is because if there is already a strong separation of flux, a fast meridional flow will lead to an enhancement of the polar field due to flux concentration.  This unrealistically strong separation is typical of kinematic dynamo models that use a non-local $\alpha$-effect BL source (see Fig.~\ref{Fig_MFvsDR}-c).  The reason is that by increasing the vector potential $A$ proportionally to the toroidal field $B$ at the bottom of the CZ (Eq.~\ref{Eq_2.5DynA}), one creates strong gradients in the vector potential above the edges of the toroidal field belt; this ends up immediately producing poloidal field which is as large in length scale as the toroidal field itself, circumventing the whole process of flux transport by circulation and diffusion.  Figure~\ref{Fig_MFvsDR} illustrates this fundamental difference: The top row shows the evolution of the surface magnetic field for a dynamo model using the double-ring algorithm (Fig.~\ref{Fig_MFvsDR}-a) versus one using the $\alpha$-effect formulation (Fig.~\ref{Fig_MFvsDR}-b).  The different way in which each formulation handles the surface dynamics is evident.  The double-ring simulation clearly shows a mixture of polarities and small-scale features which migrate to the poles (very much like the observed evolution of the surface magnetic field).  On the other hand, the mean field formulation only shows two large scale polarities whose centroids drift apart as the cycle progresses.  The bottom row depicts a snapshot of the poloidal field for the double-ring algorithm (Fig.~\ref{Fig_MFvsDR}-c) and the $\alpha$-effect formulation (Fig.~\ref{Fig_MFvsDR}-d) -- both snapshots taken at solar max.  Here the presence of small-scale features and a mixture of polarities is evident for the double ring, whereas the $\alpha$-effect formulation only shows a large-scale magnetic field with two polarities.  It is clear that although the large scale internal field is similar for both, the double-ring algorithm does a much better job of capturing the surface dynamics.

\subsection{Polar Field Strength vs. Meridional Flow Speed}

\begin{figure}
\begin{center}
  \includegraphics[scale=0.7]{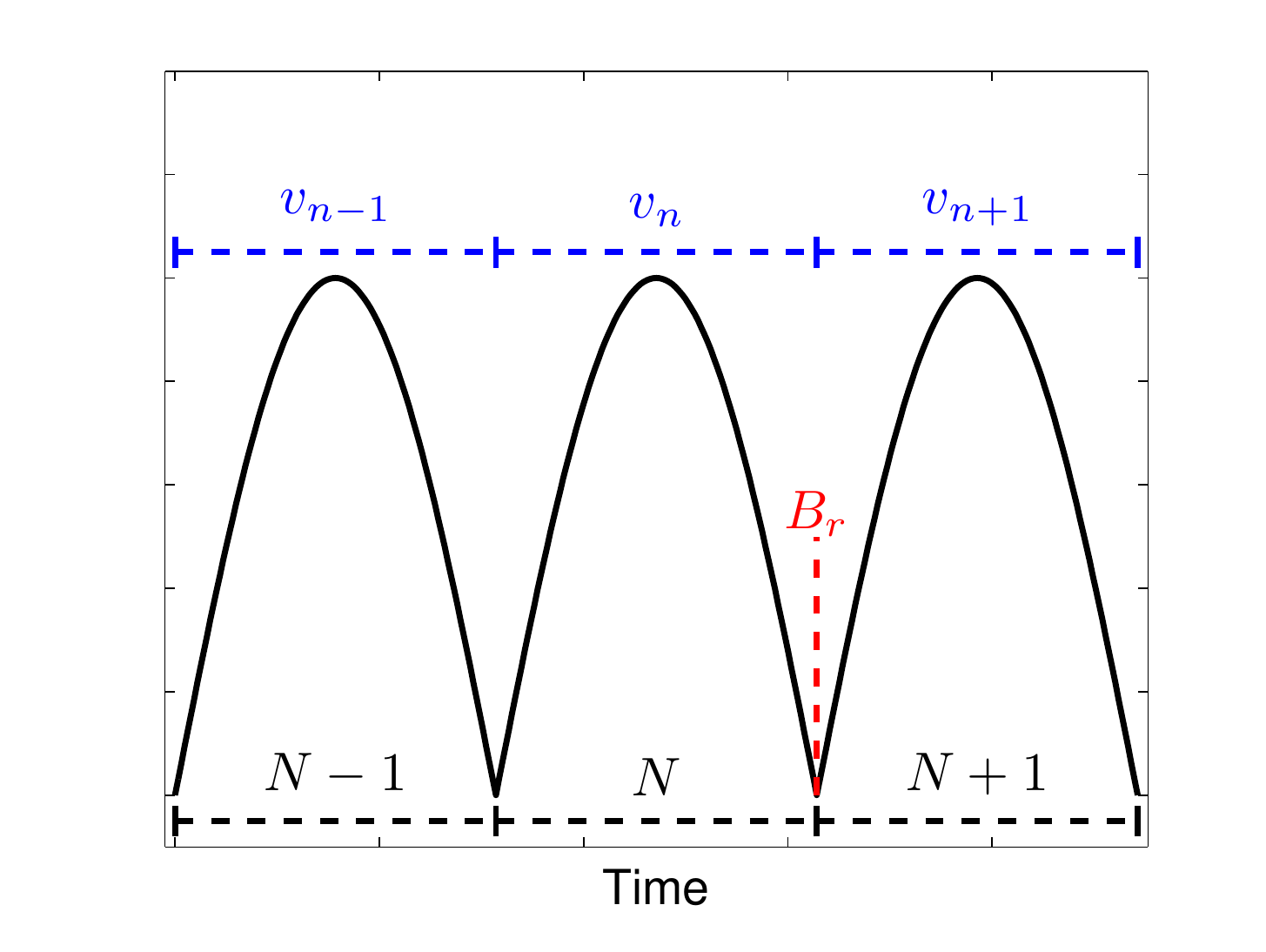}
\end{center}
\caption{Diagram for the evolution of the meridional flow amplitude with respect to the sunspot cycle: each solar cycle $N$ has a unique meridional flow strength $v_n$ which is randomly chosen between $15-30$ m/s.  Additionally, the polar field strength $B_r$ of cycle $N$ it is measured at the end of it. }\label{Fig_MFV}
\end{figure}

In order to study the relationship between meridional flow and polar field strength, we perform simulations in which we randomly change the meridional flow amplitude from one sunspot cycle to another (between $15-30$ m/s).  This is illustrated in Fig.~\ref{Fig_MFV} where a series of sunspot cycles is plotted along with their associated meridional flow.  We then evaluate the correlation between the amplitude of the meridional flow of a given cycle and the polar field strength $B_r$ at the end of it.   Since we want to evaluate the relative performance of the double-ring algorithm as opposed to the non-local BL source, we perform the same simulation for both types of sources.  Aside from the varying meridional flow amplitude and the poloidal source, the rest of the ingredients are the same.  It is important to note that partly due to difficulties in tracking the exact occurrence of solar minimum, the two hemispheres eventually drift out of phase in long simulations -- sometimes this phase difference leads to quadrupolar solutions which often go back to the observed dipolar solution.  This parity issue only appears when the meridional flow is changed at solar minimum: if there are no variations, or if the variation takes place at solar maximum, the cycle is always locked in phase with dipolar parity.  Nevertheless, to compare our simulations with surface flux-transport models, we change the flow speed only at solar minimum.  To be consistent, we accumulate statistics only from cycles in which the two hemispheres are in dipolar phase.  The statistics performed for both types of source contain about 200 sunspot cycles.

\begin{figure}
\begin{center}
\begin{tabular}{c}
             \textbf{$\alpha$-effect BL formulation}\\
    \begin{tabular}{cc}
           Scatter Plot               &   2D Histogram\\
        \includegraphics[scale=0.45]{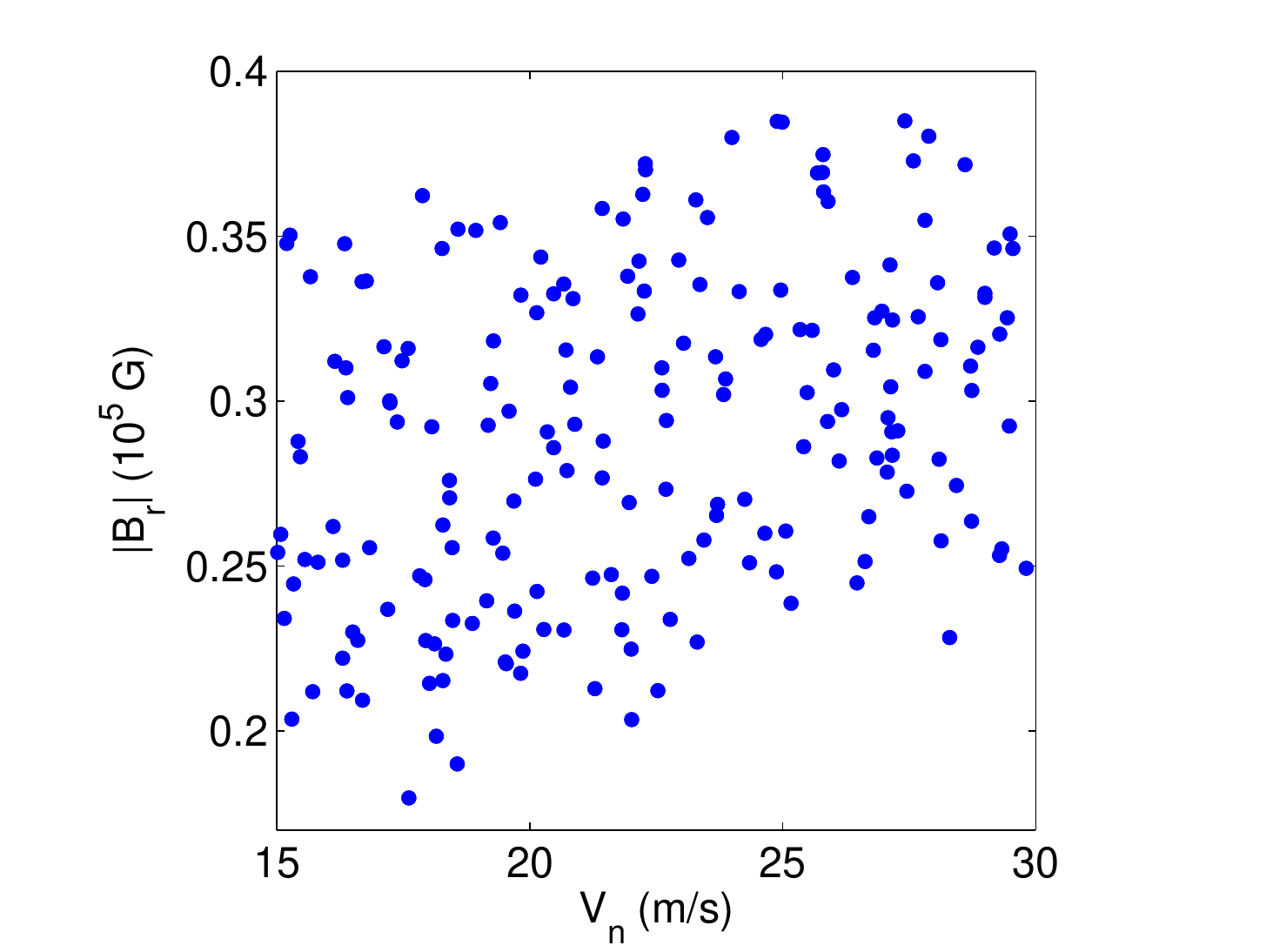} & \includegraphics[scale=0.45]{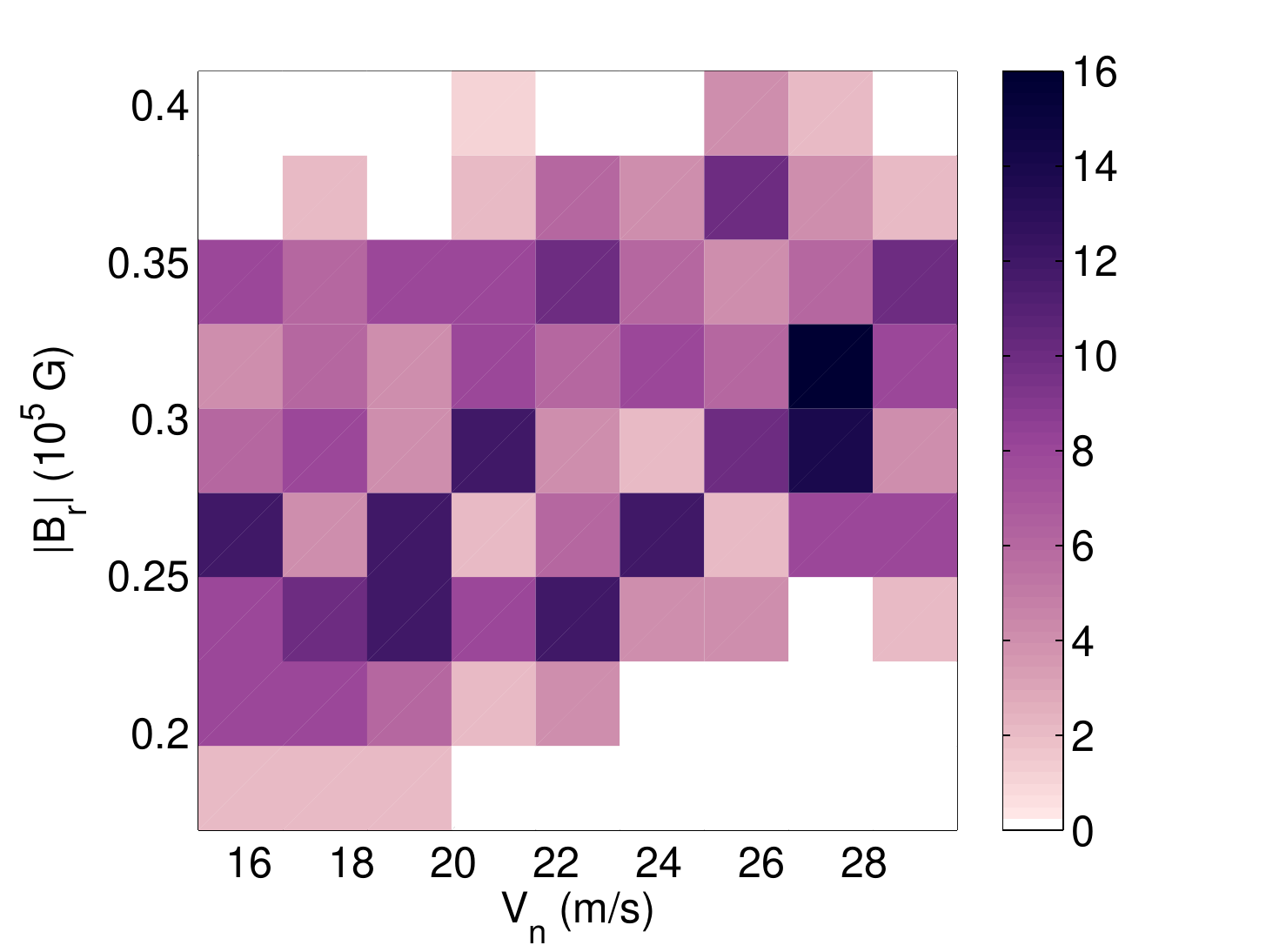}\\
                 (a)                   &                   (b)
    \end{tabular} \\
                 \textbf{Double-ring algorithm}\\
    \begin{tabular}{cc}
    Scatter Plot               &   2D Histogram\\
        \includegraphics[scale=0.45]{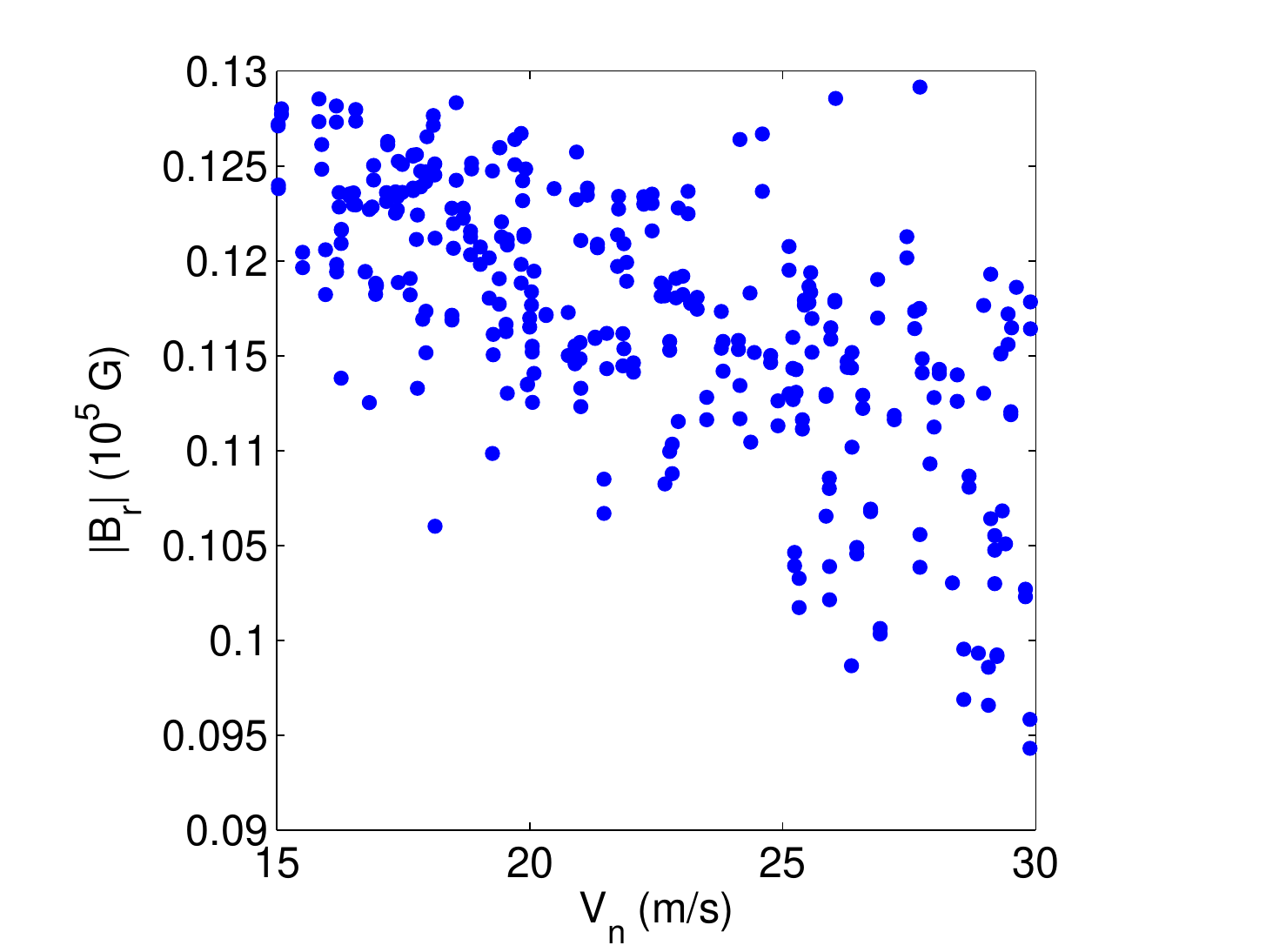} & \includegraphics[scale=0.45]{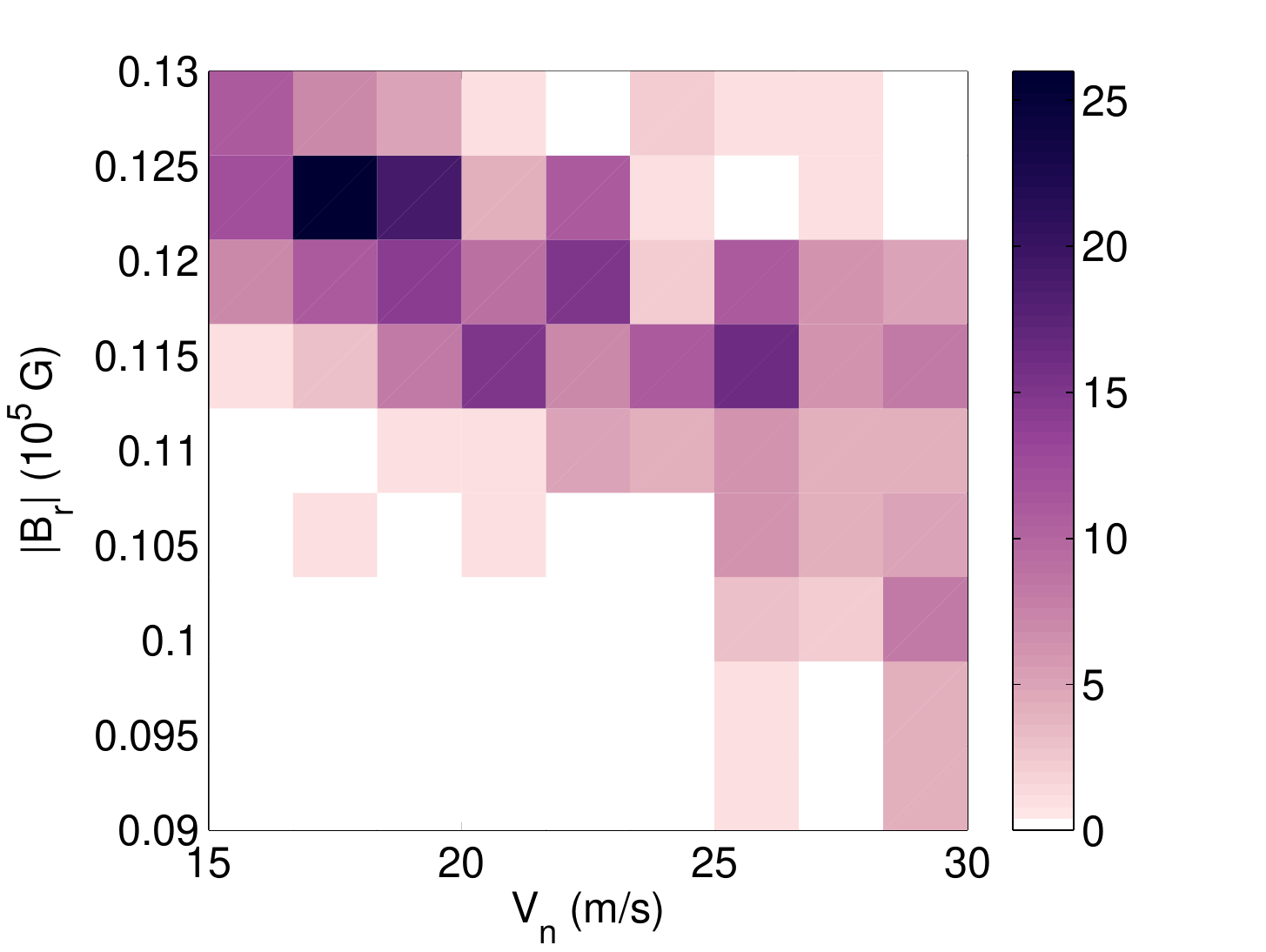}\\
                 (c)                   &                   (d)
    \end{tabular}
\end{tabular}
\end{center}
\caption{Relationship between randomly varying meridional flow speed and polar field strength for simulations using the mean-field formulation (top row) versus simulations using the double-ring algorithm (bottom-row).  The polar field strength (in Gauss) is represented by the maximum amplitude of the polar radial field ($Br$) attained during solar minimum. The relationship between the above parameters is determined by the Spearman's rank correlation coefficient. Top-row: (correlation coefficient, r     0.325, confidence, p   99.99\%). Bottom-row: (r     -0.625, p   99.99\%)}\label{Fig_MF_Obs_Ch3}
\end{figure}

The values of polar field we find using the kinematic dynamo simulations are of the order of 10 kG which is a common feature of dynamo models, which are successful in simulating the strong toroidal field necessary to produce sunspots and sustain the solar cycle (Dikpati \& Charbonneau 1999\nocite{dikpati-charbonneau99}; Chatterjee, Nandy \& Choudhuri 2004\nocite{chatterjee-nandy-choudhuri04}; Jiang \& Wang 2007\nocite{jiang-wang07}; Jouve at al.~2008\nocite{Jouve-etal08}).  Recent high resolution observations of the polar region have now confirmed the existence of such strong kilo-Gauss unipolar flux tubes (Tsuneta et al.~2008\nocite{tsuneta-etal08}).  Fig.~\ref{Fig_MF_Obs_Ch3} shows the results of both simulations.  We find a weak positive correlation between meridional flow and polar field strength for the simulations using the non-local $\alpha$-effect formulation (Fig.~\ref{Fig_MF_Obs_Ch3}-top), which is in general agreement with the results of Dikpati, de Toma \& Gilman (2008\nocite{dikpati-detoma-gilman08}).  On the other hand, the simulations using the double-ring formulation distinctively show a negative correlation (Fig.~\ref{Fig_MF_Obs_Ch3}-bottom), in agreement with surface flux-transport simulations (Wang, Sheeley \& Lean 2002\nocite{wang-sheeley-lean02}).  This clearly establishes that the discrepancy between the models is resolved by introducing the double-ring algorithm and that the double-ring formalism does a better job at capturing the observed surface magnetic field dynamics than the non-local $\alpha$-effect formalism.

\section{Concluding Remarks}

In the first half of this work, we perform surface flux-transport simulations to test the validity of the axisymmetric formulation of the kinematic dynamo problem. Our results suggest that this axisymmetric formulation captures well the surface flux dynamics over spatial and temporal scales that are relevant for the solar cycle.  Building upon this we introduce an improved version of the double-ring algorithm to model the Babcock-Leighton mechanism for poloidal field regeneration in axisymmetric, kinematic dynamo models.  We show that this new double-ring formulation generates surface field evolution and polar field reversal which is in close agreement with observations.  Additionally, we find that this improved treatment of the Babcock-Leighton process generates an inverse relationship between meridional flow speed and polar field strength -- which is suggested by simple physical arguments and also predicted by surface-flux transport simulations.  This resolves the discrepancy between kinematic dynamo models and surface flux-transport simulations regarding the dynamics of the surface magnetic field.  Since the latter drives the evolution of the corona and the heliosphere, our work opens up the possibility of coupling dynamo models of the solar cycle with coronal and heliospheric field evolution models.

\acknowledgements

\section{Acknowledgements}

We want to thank Aad van Ballegooijen for useful discussions that were crucial for the development of the algorithm mentioned in Section~\ref{sec_dblr2pld}.  The computations required for this work were performed using the resources of Montana State University and the Harvard-Smithsonian Center for Astrophysics.  We thank Keiji Yoshimura at MSU, and Alisdair Davey and Henry (Trae) Winter at the CfA for much appreciated technical support.  This research was funded by NASA Living With a Star grant NNG05GE47G and has made extensive use of NASA's Astrophysics Data System. D.N. acknowledges support from the Government of India through the Ramanujan Fellowship.  A.R.Y. thanks the UK STFC for financial support.

\bibliographystyle{apj}
%\bibliography{References}

% Figure 1

% Figure 2

%Figure 3

%Figure 4

%Figure 5

\end{document}